\documentclass[prd,aps]{revtex4}

\newcommand{\codtil}{\tilde{d}^{\dagger}}
\newcommand{\gtens}{\mbox{\boldmath $g$}}
\newcommand{\Atens}{\mbox{\boldmath $A$}}
\newcommand{\Dtens}{\mbox{\boldmath $D$}}
\newcommand{\Ftens}{\mbox{\boldmath $F$}}
\newcommand{\Ptens}{\mbox{\boldmath $P$}}
\newcommand{\Stens}{\mbox{\boldmath $S$}}
\newcommand{\gtiltens}{\tilde{\gtens}}
\newcommand{\ghattens}{\hat{\gtens}}
\newcommand{\gtil}{\tilde{g}}
\newcommand{\ghat}{\hat{g}}
\newcommand{\laptil}{\tilde{\Delta}}
\newcommand{\nabtil}{\tilde{\nabla}}
\newcommand{\asttil}{\tilde{\ast}}
\newcommand{\etatil}{\tilde{\eta}}
\newcommand{\laphat}{\hat{\Delta}}
\newcommand{\nabhat}{\hat{\nabla}}
\newcommand{\asthat}{\hat{\ast}}

\begin{document}

\title{Stability properties of black holes in self-gravitating nonlinear electrodynamics}

\author{Claudia Moreno}
\affiliation{Center for Gravitational Physics and Geometry,
             Department of Physics, The Pennsylvania State University,
             104 Davey Laboratory, University Park, Pennsylvania 16802}
\affiliation{Department of Physics and Astronomy,
             The University of Texas at Brownsville,
             80 Fort Brown, Brownsville, Texas 78520--4993}

\author{Olivier Sarbach}
\affiliation{Department of Physics and Astronomy,
             Louisiana State University,
             202 Nicholson Hall, Baton Rouge, Louisiana 70803--4001}

\date{\today}

\begin{abstract}
We analyze the dynamical stability of black hole solutions in self-gravitating
nonlinear electrodynamics with respect to arbitrary linear fluctuations
of the metric and the electromagnetic field. In particular, we derive
simple conditions on the electromagnetic Lagrangian which imply
linear stability in the domain of outer communication. 
We show that these conditions hold for several of the regular black hole 
solutions found by Ay\'on-Beato and Garc\'{\i}a.
\end{abstract}

\maketitle

\section{Introduction}

In the last few years, interesting exact solutions 
\cite{ABG-E1,ABG-E2,ABG-E3,ABG-M1,Br,BH,EH}
have been obtained in the context of nonlinear electrodynamics
(NED) coupled to gravity. These solutions are interesting 
because for a suitable choice of the electromagnetic
Lagrangian they describe static black hole spacetimes
where the black hole does not exhibit the usual singularity
at its center, but instead all fields are everywhere regular 
and fall off in the asymptotic region. 
The existence of these solutions does not contradict the 
singularity theorems \cite{HE} since the corresponding spacetimes 
are not globally hyperbolic and their stress-energy tensor does not 
satisfy the strong energy condition.
Examples of regular black hole spacetimes have been given by
Bardeen \cite{Bardeen}. While the corresponding Einstein tensor
satisfies the weak energy condition these spacetimes do not
satisfy the vacuum Einstein equations. As shown recently by
Ay\'on-Beato and Garc\'{\i}a \cite{ABG-M1}
the non-vanishing Einstein tensor in the Bardeen model can be
associated with the stress-energy tensor of a nonlinear
electromagnetic Lagrangian and the Bardeen black holes can be 
obtained as exact solutions of an appropriate NED coupled to gravity.

An important issue in field theories is the study of stability
of a given static or stationary field configuration. Unstable
solutions have usually less physical significance than stable ones
since they are likely to decay to a stable configuration.
For example, it is known that static spherically symmetric regular
particle-like solutions in Einstein-Yang-Mills theories 
are linearly unstable and numerical studies suggest that 
(depending on the details of the initial perturbations)
a perturbation either causes the system to explode or to
collapse to a Schwarzschild black hole (which is linearly stable)
\cite{Zhou-Diss}.

In the context of regular black hole solutions in NED, we
study the behavior of linear (classical) fluctuations of the
gravitational and electromagnetic fields in the domain of
outer communication. Basically, there are two possible outcomes:
1) The fluctuations grow exponentially in time which means
that the background configuration is (at least linearly) unstable.
In this case a natural question to ask is to what configuration
(if any) the solution will decay to.
2) The fluctuations are bounded in time which means that the
background configuration is linearly stable.
This suggests that the perturbed configuration
eventually settles down to the background
configuration after the possible emission of gravitational
and electromagnetic radiation.

In this article, we derive simple conditions on the 
electromagnetic Lagrangian that guarantee 
the corresponding black hole solutions to be linearly stable. 
In order to do so we cast the perturbation equations into a system 
of coupled wave equations with symmetric potential. Linear stability
can then be established if this potential is positive 
definite. We show that the latter condition holds if
the electromagnetic Lagrangian and its first and second 
order derivatives satisfy appropriate inequalities.
We check these inequalities for several of the regular black holes 
that have been found in self-gravitating NED and so prove that
these solutions are linearly stable.
However, we also argue that this situation is not generic.
That is, one can easily obtain an unstable regular black hole
solution by slightly changing the electromagnetic Lagrangian 
belonging to a regular black hole solution.

A brief review of the theory of NED in general relativity 
and of the regular black hole solutions
is presented in section \ref{Sect-2}. 
In section \ref{Sect-3} we derive the perturbation equations
and cast them into the form of a wave equation with potential.
We derive sufficient conditions for this potential to be positive
and show in section \ref{Sect-4} that the latter condition 
implies the linear stability of the black holes.
Conclusions are drawn in section \ref{Sect-5}.
A demonstration of the fact that the Bardeen black holes are
linearly stable is given in an appendix at the end of this 
article.

\section{Regular black holes in self-gravitating nonlinear electrodynamics}
\label{Sect-2}

In this section, we first review a few facts from NED
and then focus on spherically symmetric solutions. 
In particular, we give a brief description of
the regular black hole solutions obtained in
\cite{ABG-E1, ABG-E2, ABG-E3}.

\subsection{Equations of motion}

The action for NED in general relativity is
\begin{displaymath}
S\left[ \gtens,\Atens \right] = \frac{1}{4\pi}\int\sqrt{-g}\, d^4 x 
\left( \frac{1}{4} R - {\cal L}(F) \right),
\end{displaymath}
where $R$ denotes the Ricci scalar with respect to the spacetime
metric $g_{\mu\nu}$, $F_{\mu\nu} = \partial_\mu A_\nu - \partial_\nu A_\mu$
is the electromagnetic field, and ${\cal L}$ is an arbitrary
function in the invariant $F\equiv \frac{1}{4} F^{\mu\nu} F_{\mu\nu}$.
In Einstein-Maxwell theory, ${\cal L}(F) = F$, but here we
consider more general choices of electromagnetic 
Lagrangians.\footnote{However we do not consider the case where ${\cal L}$
depends on the invariant $F_{\mu\nu}\ast F^{\mu\nu}$ as well.}
The field equations are
\begin{eqnarray}
&& G^\mu_{\;\nu} = 2\left[ {\cal L}_F F^{\mu\alpha} F_{\nu\alpha} 
- \delta^\mu_{\;\nu}{\cal L} \right],
\label{Eq:EinsteinF}\\
&& \nabla_\mu\left( {\cal L}_F F^{\alpha\mu} \right) = 0,
\label{Eq:NMaxF}\\
&& \nabla_\mu \ast F^{\alpha\mu} = 0,
\label{Eq:NBianchiF}
\end{eqnarray}
where ${\cal L}_F \equiv \partial {\cal L}/\partial F$ and 
$G^\mu_{\;\nu}$, $\nabla$ and $\ast$ denote the components of the 
Einstein tensor, the covariant derivative and the Hodge dual with 
respect to $g_{\mu\nu}$, respectively.\footnote{We use the metric 
signature $(-,+,+,+)$ throughout, and Greek letters
denote spacetime indices taking values $(0,1,2,3)$. 
The Hodge dual of $F_{\mu\nu}$ is given by
$\ast F^{\mu\nu} = \frac{1}{2} \eta^{\mu\nu\alpha\beta} F_{\alpha\beta}$,
where $\eta_{\mu\nu\alpha\beta}$ is completely skew-symmetric
with $\eta_{0123} = \sqrt{-g}$.}
We have included the Bianchi identities (\ref{Eq:NBianchiF})
for completeness and later use.

It is also convenient to express the field equations in terms of the
field $P_{\mu\nu} \equiv {\cal L}_F F_{\mu\nu}$ and the Hamiltonian
${\cal H} \equiv 2{\cal L}_F F - {\cal L}$ which we consider to
be a function of $P \equiv \frac{1}{4} P^{\mu\nu} P_{\mu\nu}$ \cite{SGP}:
\begin{eqnarray}
&& G^\mu_{\;\nu} = 2\left[ {\cal H}_P P^{\mu\alpha} P_{\nu\alpha} 
- \delta^\mu_{\;\nu}\left( 2{\cal H}_P P - {\cal H} \right) \right],
\label{Eq:Einstein}\\
&& \nabla_\mu P^{\alpha\mu} = 0,
\label{Eq:NMax}\\
&& \nabla_\mu\left( \ast {\cal H}_P P^{\alpha\mu} \right) = 0.
\label{Eq:NBianchi}
\end{eqnarray}
However, one should notice that the set of field equations
(\ref{Eq:EinsteinF},\ref{Eq:NMaxF},\ref{Eq:NBianchiF}) (called the $F$ framework)
are equivalent to the set of equations 
(\ref{Eq:Einstein},\ref{Eq:NMax},\ref{Eq:NBianchi}) (called the $P$ framework) 
only if the map $F \mapsto P$ is invertible. 
Since $P = {\cal L}_F^2 F$, this requires that
\begin{eqnarray}
0 \neq \frac{\partial P}{\partial F} &=& 
  {\cal L}_F\left( {\cal L}_F + 2{\cal L}_{FF} F \right),
\quad\hbox{and}\nonumber\\
0 \neq \frac{\partial F}{\partial P} &=& 
  {\cal H}_P\left( {\cal H}_P + 2{\cal H}_{PP} P \right).
\label{Eq:FPEquiv}
\end{eqnarray}

Finally, we mention that there is a duality between the
$F$ framework and the $P$ framework:
Using the fact that
\begin{displaymath}
{\cal L}_F {\cal H}_P = 1, \qquad
{\cal L} = 2{\cal H}_P P - {\cal H}, \qquad
\ast P^{\mu\alpha} \ast P_{\nu\alpha} = P^{\mu\alpha} P_{\nu\alpha} - 2\delta^\mu_{\;\nu} P,
\end{displaymath}
it is not difficult to see that the equations 
(\ref{Eq:Einstein},\ref{Eq:NMax},\ref{Eq:NBianchi})
can be obtained from the equations 
(\ref{Eq:EinsteinF},\ref{Eq:NMaxF},\ref{Eq:NBianchiF}) by
performing the following transformation \cite{Br}
\begin{eqnarray}
g_{\mu\nu} &\mapsto& \bar{g}_{\mu\nu} \equiv g_{\mu\nu}\, \nonumber\\
F_{\mu\nu} &\mapsto& \bar{F}_{\mu\nu} \equiv \ast P_{\mu\nu}\, ,
\label{Eq:FPDuality}\\
{\cal L}(F) &\mapsto& \bar{{\cal L}}(\bar{F}) \equiv -{\cal H}(P) = -{\cal H}(-\bar{F}).
\nonumber
\end{eqnarray}
Notice however that the $FP$ duality transformation 
(\ref{Eq:FPDuality}) is only a {\em symmetry} of the system if 
$\bar{{\cal L}}(F) = {\cal L}(F)$, i.e. if
${\cal L}(-F) + 2F {\cal L}_{F}(-F) = {\cal L}(F)$.
The only Lagrangian that satisfies this identity is
${\cal L}(F) = \alpha F + \Lambda/2$, with $\alpha$ and
$\Lambda$ two constants. If $\alpha=1$ this is the Maxwell
Lagrangian with a cosmological term.
In general, the $FP$ duality connects theories
belonging to {\em different} Lagrangians.
(For a more detailed analysis of duality invariant 
nonlinear electrodynamics, see Refs. \cite{SGP, GR}.)

\subsection{Spherically symmetric solutions}

We now focus on spherically symmetric and purely electric
solutions in the $P$ framework, equations
(\ref{Eq:Einstein},\ref{Eq:NMax},\ref{Eq:NBianchi}). 
{}From this, one can also obtain purely magnetic solution in 
the $F$ framework (but with a different Lagrangian) by 
applying the inverse of the $FP$ duality transformation 
(\ref{Eq:FPDuality}).
In the remainder of this section, we assume that the
Hamiltonian ${\cal H}(P)$ is negative and satisfies the 
{\em weak field limit}
\begin{displaymath}
{\cal H}(P) \simeq P, \;\; \hbox{as} \;\; P \rightarrow 0,
\end{displaymath}
although this condition will not be a crucial assumption 
in the stability analysis below.

A spherically symmetric spacetime can locally
be written as $\tilde{M} \times S^2$ where $\tilde{M}$ is a
two-dimensional pseudo-Riemannian manifold. The metric takes
the form
\begin{displaymath}
\gtens = \gtil_{ab}\, dx^a dx^b + r^2 d\Omega^2, 
\end{displaymath}
where $d\Omega^2 = d\vartheta^2 + \sin^2\vartheta\, d\varphi^2$ 
is the standard metric on the two-sphere $S^2$ and 
$\gtiltens = \gtil_{ab}\, dx^a dx^b$ 
is a metric of signature $(-1,1)$, where the indices $a,b$ label
some coordinates on the manifold $\tilde{M}$.
It is not difficult to see that
\begin{equation}
P_{ab} = \frac{Q}{r^2}\, \etatil_{ab}\; ,
\label{Eq:Ptensback}
\end{equation}
with the other components of $P_{\mu\nu}$ being zero,
satisfies the NED equations (\ref{Eq:NMax},\ref{Eq:NBianchi}).
Here, $\etatil_{ab}$ is the volume element corresponding to $\gtiltens$
and $Q$ is a constant that represents the electric charge.

Next, Einstein's field equations (\ref{Eq:Einstein}) reduce to
\begin{eqnarray}
G_{ab} &=& 2\gtil_{ab} {\cal H}, \nonumber\\
G_{Ab} &=& 0, \nonumber\\
G_{AB} &=& -2 g_{AB} {\cal L}, \nonumber
\end{eqnarray}
where capital indices refer to the angular variables $\vartheta$ and $\varphi$.
In Schwarzschild coordinates,
\begin{equation}
\gtiltens = -S^2\, N\, dt^{\, 2} +\frac{dr^{\, 2}}{N}\; ,
\label{Eq:SchwCoord}
\end{equation}
these equations yield $S = S(t)$ (which can be set to $1$ by an
appropriate redefinition of $t$) and
\begin{equation}
\partial_t m = 0, \qquad
\partial_r m = -r^2 {\cal H}(P), \qquad
P = -\frac{1}{2}\frac{Q^2}{r^4}\, ,
\label{Eq:mprime}
\end{equation}
where the mass function $m$ is defined by $N = 1 - 2m/r$.
In particular, it follows from (\ref{Eq:mprime}) that spacetime
is static (as long as $N > 0$), i.e. Birkhoff's theorem \cite{HE}
holds.

Given the Hamiltonian ${\cal H}(P)$, equations (\ref{Eq:mprime})
can be integrated to yield the mass function $m(r)$:
\begin{equation}
m(r) = M + \int_r^\infty r^2 {\cal H}\, d r,
\label{Eq:mass}
\end{equation}
where $M$ is an integration constant representing the ADM mass.
As a consequence of the weak field limit, we have
$m(r) \rightarrow M - Q^2/2r$ as $r\rightarrow\infty$
and the metric is asymptotically flat 
with the same asymptotic as the Reissner-Nordstr\"om metric.
If the Hamiltonian ${\cal H}(P)$ is such that the integral
in equation (\ref{Eq:mass}) converges as $r \rightarrow 0$, we can
choose the ADM mass such that $m(r = 0) = 0$. In particular if
${\cal H} \rightarrow -\Lambda/2$ as $r \rightarrow 0$ where 
$\Lambda$ is a positive constant, one obtains 
$m(r) \rightarrow \Lambda r^3/6$, that is, the metric approaches 
the one of de Sitter space as $r$ approaches zero.
Hamiltonians with a similar type of behavior at $r=0$ and $r=\infty$ 
have been constructed in Ref. \cite{ABG-E1, ABG-E2, ABG-E3}.
As an explicit example, we consider the one that has been used
in Ref. \cite{ABG-E1}:
\begin{equation}
{\cal H}(P) = P\frac{1 - 3x}{(1+x)^3} - \frac{3}{2Q^2 s}\left( \frac{x}{1+x} \right)^{5/2},
\label{Eq:ABG-E1Ham}
\end{equation}
where $x = \sqrt{-2Q^2P} = Q^2/r^2$ and $s = |Q|/2m$.
The corresponding mass function is obtained from (\ref{Eq:mass}):
\begin{displaymath}
m(r) = M -\frac{1}{2}\frac{ Q^2 r^3}{(r^2+Q^2)^2} + m\frac{r^3}{(r^2+Q^2)^{3/2}} - m.
\end{displaymath}
Thus, if $M = m$, the metric is regular at $r=0$ and so
is the only non-trivial component of the electric field 
$E = F_{tr} = {\cal H}_P P_{tr}$:
\begin{displaymath}
E = Qr^4\left( \frac{r^2 - 5Q^2}{(r^2 + Q^2)^4} + \frac{15}{2}\frac{M}{(r^2+Q^2)^{7/2}} \right).
\end{displaymath}
Furthermore, if the charge to mass ratio $(2s)$ is small enough,
the function $N(r)$ intersects the $r$-axis at some values
$r_c$, $r_h$ with $r_c < r_h$, $r_h$ and $r_c$ representing
the radius of an event and Cauchy horizon, respectively.
Properties of the Penrose diagrams belonging to these solutions
have been discussed in \cite{ABG-E1}. As expected, the 
corresponding spacetimes possess the same global structure as 
the Reissner-Nordstr\"om black holes except that there is no 
singularity at $r=0$; $r=0$ is the origin of the spherical 
coordinates.

We end this section after some comments.
First, it is clear that there is a wide class of Hamiltonians 
that have the qualitative properties described above, since one 
could deform slightly the Hamiltonian (\ref{Eq:ABG-E1Ham}) without 
affecting its asymptotic properties.
Next, as observed in \cite{Br, Br-Lett}, regular electrically
charged black holes are not well-defined when interpreted in the
$F$ framework. What happens is that since $F = -E^2/2$ is negative 
but vanishes as $P$ goes to zero or infinity, it must exhibit 
a minimum at some $P^*$. Therefore, the condition (\ref{Eq:FPEquiv})
is violated at at least one point and the electromagnetic
Lagrangian ${\cal L}(F)$ suffers from branching, see Ref. \cite{Br}.
Next, while the stress-energy tensor corresponding to
(\ref{Eq:ABG-E1Ham}) satisfies the {\em weak} energy condition
(because ${\cal H} \leq 0$) it violates the {\em strong} energy condition
since spacetime approaches the one of de Sitter near the
horizon. This, together with the fact that the spacetimes
possess Cauchy horizons, explains why there is no contradiction
to the singularity theorems.
Finally, magnetically charged solutions can also be obtained
by using the $FP$ duality transformation (\ref{Eq:FPDuality}),
as mentioned above.
An example of a magnetically charged solution is provided by the
Bardeen model which has been interpreted as a self-gravitating 
nonlinear magnetic monopole in \cite{ABG-M1}.

\section{Perturbation equations}
\label{Sect-3}

In this section we investigate the stability of the
solutions described in section \ref{Sect-2} with respect to 
non-spherically symmetric dynamical linear fluctuations 
outside the event horizon.
In order to do so, we use a recently introduced gauge-invariant
perturbation formalism \cite{S-Diss, ST} and generalize the 
equations obtained by Moncrief \cite{Moncrief} to nonlinear 
electromagnetic fields.
These equations (which we shall call the pulsation equations)
are a set of wave-like coupled equations with symmetric potential
acting on gauge-invariant
and unconstrained quantities which therefore capture the purely
physical degrees of freedom associated to the linear fluctuations.
Our perturbation formalism, which is described in more details in \cite{ST} 
for vacuum perturbations and in \cite{S-Diss} for the Einstein-Maxwell
case, has the additional advantage to yield the
pulsation equations for any spherically symmetric coordinates
on the background. 

Notice that an alternative approach is given by the Newman-Penrose
formalism (the background has a metric of type D). However, we
have not succeeded in decoupling the equations along the 
lines of \cite{C-Book}. As we will see at the end of this section,
one does not necessarily have the equivalence between the spectra
belonging to the pulsation operators in the odd- and even-parity
sectors when the electromagnetic Lagrangian is nonlinear.
This might be an indication that the Newman-Penrose approach is
difficult when nonlinear matter fields are coupled to the metric.

The reader who is not interested in the derivation of the
pulsation equations may skip the remaining of this section.

\subsection{Preliminaries}

Our task is to linearize the equations 
(\ref{Eq:Einstein},\ref{Eq:NMax},\ref{Eq:NBianchi})
around a spherically symmetric background, where we have to keep
in mind that 
$P_{\mu\nu} = {\cal H}_P^{-1} (\partial_\mu A_\nu - \partial_\nu A_\mu)$.
In the following, we assume that the function ${\cal H}_P$ 
is regular and positive for all $r > r_h$, where $r_h$ denotes
the radius of the event horizon.

In terms of small perturbations of the metric and the vector potential,
$\delta g_{\mu\nu}$ and $\delta A_\mu$, respectively, the linearized
equations become
\begin{eqnarray}
\delta G_{\mu\nu} &=& 2\left\{ {\cal H}_{PP}  P_\mu^{\;\alpha} P_{\nu\alpha}\, \delta P 
 + {\cal H}_P \left( 2P_{(\mu}^{\;\;\;\alpha}\delta P_{\nu)\alpha} 
 - P_\mu^{\;\sigma} P_\nu^{\;\rho}\delta g_{\sigma\rho} \right) \right. 
\nonumber\\
&& \qquad \left. -\, g_{\mu\nu} \kappa {\cal H}_P \delta P
    - \delta g_{\mu\nu}\left( 2{\cal H}_P P - {\cal H} \right) \right\},
\label{Eq:LinEinstein}\\
0 &=& \nabla^\mu\left( \delta P_{\alpha\mu} - 2\delta g_{\sigma [\alpha} P^\sigma_{\; \mu]}
+ \frac{1}{2}\, g^{\sigma\rho}\delta g_{\sigma\rho}\, P_{\alpha\mu} \right),
\label{Eq:LinNMax}
\end{eqnarray}
where in the following the function
\begin{equation}
\kappa = 1 + 2{\cal H}_P^{-1} {\cal H}_{PP} P
\label{Eq:KDef}
\end{equation}
will play an important role. Notice that in view of
the inequality (\ref{Eq:FPEquiv}) and our assumption on 
${\cal H}_P$ to be regular and positive, $\kappa$ is regular and
non-zero outside the event horizon if and only if the portion 
of the background solution that lies outside the event horizon 
is well-defined in the $F$ framework.

The variation of $P_{\mu\nu}$ can be expressed as
\begin{equation}
\delta P_{\mu\nu} = \frac{1}{2}\,{\cal H}_P^2 {\cal L}_{FF}
\left( P^{\alpha\beta}\delta F_{\alpha\beta} 
 - {\cal H}_P P^{\sigma\alpha} P^\rho_{\;\alpha} \delta g_{\sigma\rho} \right) P_{\mu\nu} 
 + {\cal H}_P^{-1} \delta F_{\mu\nu},
\label{Eq:LinPtens}
\end{equation}
where in terms of ${\cal H}(P)$, we have
\begin{displaymath}
{\cal L}_{FF} = -\kappa^{-1} {\cal H}_P^{-4} {\cal H}_{PP}\, .
\end{displaymath}
General expressions for the perturbed Einstein tensor are found,
for instance, in \cite{S-Diss}.

The perturbed quantities $\delta g_{\mu\nu}$ and
$\delta A_\mu$ are subject to the following gauge transformations:
With respect to an infinitesimal coordinate transformation
$\delta x^\mu \mapsto \delta x^\mu + X^\mu$, where $X^\mu$ is
a vector field,
\begin{equation}
\delta g_{\mu\nu} \mapsto \delta g_{\mu\nu} + \pounds_X g_{\mu\nu}\; ,
\qquad
\delta A_\mu \mapsto \delta A_\mu + \pounds_X A_\mu\; ,
\label{Eq:CoordTrans}
\end{equation}
where $\pounds_X$ denotes the Lie derivative with respect to $X^\mu$.
Furthermore, the potential $A_\mu$ is subject to $U(1)$
gauge transformations generated by a scalar $\chi$:
\begin{equation}
\delta A_\mu \mapsto \delta A_\mu + \partial_\mu\chi\, .
\label{Eq:U1Trans}
\end{equation}
Because of the gauge invariance of the theory, the perturbation
equations are invariant with respect to the transformations
(\ref{Eq:CoordTrans},\ref{Eq:U1Trans}). Since we want to 
separate these transformations from the purely physical degrees 
of freedom, our strategy is to introduce gauge-invariant combinations
of the perturbation amplitudes and to recast all equations in terms
of these amplitudes.

Since the background is spherically symmetric, it is convenient to
expand all perturbations in spherical tensor harmonics. Perturbations
belonging to different angular momentum number $\ell$ and $m$ and
with different parities decouple from each other. We discuss only
the sectors with $\ell\geq 1$ since the spherically symmetric sector
($\ell=0$) has already been discussed in the previous section.

\subsection{Odd-parity sector}

In each sector $\ell m$ linear perturbations with odd parity 
are parametrized by two scalar fields $k$ and $\nu$ and a
one-form $h = h_a d x^a$ on the two-dimensional manifold $\tilde{M}$:
\begin{eqnarray}
&& \delta g_{ab} = 0, \qquad
   \delta g_{Ab} = h_b\, S_A\, ,\qquad
   \delta g_{AB} = 2k \nabhat_{(A} S_{B)}\, ,\nonumber\\
&& \delta A_\mu dx^\mu = \nu\, S_B dx^B, \nonumber
\end{eqnarray}
where in terms of the standard spherical harmonics $Y = Y^{\ell m}$
the transverse vector harmonics $S_B$ are defined by
$S_B = (\asthat dY)_B = \hat{\eta}^A_{\; B}\nabhat_A Y$.
Here and in the following, quantities with a hat refer to the
standard metric $\ghattens = d\Omega^2$ on $S^2$ and
quantities with a tilde refer to the metric $\gtiltens$
on $\tilde{M}$.
It is not difficult to show that $\nu$ is invariant
with respect to both coordinate and $U(1)$ gauge transformations.
In the gravitational sector, we can construct the gauge-invariant
one-form (see \cite{S-Diss, ST})
\begin{equation}
h^{(inv)} = h - r^2d\left( \frac{k}{r^2} \right).
\label{Eq:hinv}
\end{equation}
Since the perturbation equations are gauge invariant we can choose
a gauge in which $k=0$ and $h^{(inv)} = h$ to simplify the
calculations and then replace $h$ by $h^{(inv)} = h - r^2d(r^{-2} k)$
in the final result.

\subsubsection{Gravito-electromagnetic equations}

Since in the odd-parity sector all scalar perturbations vanish,
the expressions (\ref{Eq:LinEinstein},\ref{Eq:LinNMax},\ref{Eq:LinPtens})
simplify considerably in this case.
For example we have $\delta P = 0$, and so
$\delta P_{\mu\nu} =  {\cal H}_P^{-1} (\partial_\mu \delta A_\nu - \partial_\nu \delta A_\mu)$.
Explicitly,
\begin{equation}
\delta P_{\mu\nu} = {\cal H}_P^{-1}
\left[ d\nu\wedge S_B dx^B - \ell(\ell+1)\nu\, Y d\Omega \right]_{\mu\nu},
\label{Eq:LinPtens_odd}
\end{equation}
where we have used $d(S_B dx^B) = \laphat Y d\Omega = -\ell(\ell+1) Y d\Omega$.

Introducing equation (\ref{Eq:LinPtens_odd}) into formula (\ref{Eq:LinNMax}), 
using the background expressions (\ref{Eq:Ptensback}) and the fact that 
$g^{\alpha\beta} \delta g_{\alpha\beta} = 0$, 
the linearized electromagnetic equations yield
\begin{displaymath}
\codtil\left( {\cal H}_P^{-1} d \nu \right) + \frac{\ell(\ell+1)}{r^2} {\cal H}_P^{-1} \nu + Q \asttil d\left( \frac{h}{r^2} \right) = 0,
\end{displaymath}
where $\asttil$ and $\codtil \equiv \asttil d\asttil$ denote
the Hodge dual and the co-differential, respectively, with respect to 
the metric $\gtiltens$.\footnote{In a coordinate-dependent notation,
one has $\codtil d\nu = -\nabtil^a\nabtil_a \nu$, 
$\asttil dh = \etatil^{ab}\nabtil_a h_b$, 
$(\codtil dh)_a = 2\nabtil^b\nabtil_{[a} h_{b]}$,
$\asttil d\nu_a = \etatil^b_{\; a} \nabtil_b\nu$, etc.
}
Next, from expressions (\ref{Eq:LinEinstein}) and (\ref{Eq:LinPtens_odd})
we find
\begin{displaymath}
\delta G_{Ab} d x^b = 
4\left\{ Q (\asttil d\nu) - r^2{\cal L} h \right\}\, \frac{S_A}{2r^2}\, ,
\qquad
\delta G_{ab} = 0, \qquad
\delta G_{AB} = 0.
\end{displaymath}
Using this in connection with the corresponding expressions for the
perturbed Einstein tensor in the odd parity sector (see Ref. \cite{S-Diss}),
\begin{eqnarray}
\delta G_{Ab} d x^b &=& 
\left\{ \codtil\left[ r^4d\left(\frac{h}{r^2} \right) \right]
 + \left( \ell(\ell+1) - 2 + r^2 G^B_B\right)h \right\} \frac{S_A}{2r^2}\; ,
\nonumber\\
\delta G_{AB} &=& -\codtil h\, \nabhat_{(A} S_{B)}\; ,
\end{eqnarray}
we end up with the manifestly gauge-invariant equations
\begin{eqnarray}
\codtil\left[ r^4d\left( \frac{h^{(inv)}}{r^2} \right) \right]
 + \lambda h^{(inv)} &=& 4Q \asttil d\nu, 
\label{Eq:Odd1}\\
\codtil\left[ {\cal H}_P^{-1} d\nu \right] 
 + \frac{\ell(\ell+1)}{r^2} {\cal H}_P^{-1}\nu &=& 
 - Q\asttil d\left(\frac{h^{(inv)}}{r^2} \right), 
\label{Eq:Odd2}\\
\codtil h^{(inv)} &=& 0,
\label{Eq:Odd3}
\end{eqnarray}
where here and in the following
$\lambda \equiv (\ell-1)(\ell+2)$.
Equation (\ref{Eq:Odd3}) is void if $\ell=1$ since in this
case, $\nabhat_{(A} S_{B)} = 0$.

\subsubsection{Pulsation equations ($\ell\geq 2$)}

We first consider the case $\ell \geq 2$.
Then, equation (\ref{Eq:Odd3}) allows to express $h^{(inv)}$
in terms of a potential $\Psi$,
\begin{displaymath}
h^{(inv)} = \frac{1}{\sqrt{\lambda}}\,\asttil d(r\Psi),
\end{displaymath}
where the factors $1/\sqrt{\lambda}$ and $r$ turn out to be convenient.
Introducing also $\Phi = \sqrt{4 {\cal H}_P^{-1}}\, \nu$, we can
integrate equation (\ref{Eq:Odd1}) and eventually obtain the
following pulsation equations\footnote{When integrating equation
(\ref{Eq:Odd1}), an integration constant is set to zero by
appropriately choosing the free constant in the potential $\Psi$.}:
\begin{eqnarray}
&& -\laptil \Psi 
 + \left[ \frac{\lambda}{r^2} + r\laptil\left( \frac{1}{r} \right) \right] \Psi
 - \frac{ \sqrt{4\lambda {\cal H}_P}\, Q}{r^3}\, \Phi = 0, 
\nonumber\\
&& -\laptil \Phi
 - \frac{ \sqrt{4\lambda {\cal H}_P}\, Q}{r^3}\, \Psi
 + \left[ \frac{\ell(\ell+1)}{r^2} + {\cal H}_P^{1/2}\laptil {\cal H}_P^{-1/2}
    + \frac{4 Q^2}{r^4}\, {\cal H}_P \right] \Phi = 0,
\nonumber
\end{eqnarray}
where $\laptil \equiv -\codtil d = \nabtil^a\nabtil_a$ is the Laplace
operator with respect to the metric $\gtiltens$.
These equations have the form of a coupled system of wave equations
with symmetric potential.
We can rewrite them into the form
\begin{equation}
\left( \Dtens + \Stens \right)
\left( \begin{array}{c} \Psi \\ \Phi \end{array} \right) = 0, 
\label{Eq:PulsOdd}
\end{equation}
where
\begin{eqnarray}
\Dtens &=& \left( \begin{array}{cc} 
 -r\nabtil^a \frac{1}{r^2} \nabtil_a r & 0 \\
 0 & -{\cal H}_P^{1/2} \nabtil^a {\cal H}_P^{-1} \nabtil_a {\cal H}_P^{1/2} 
\end{array} \right), \nonumber\\
\Stens &=& \frac{1}{r^2} \left( \begin{array}{cc} 
 \lambda & -\frac{\sqrt{4\lambda {\cal H}_P}\, Q}{r} \\
 -\frac{\sqrt{4\lambda {\cal H}_P}\, Q}{r} & \ell(\ell+1) + \frac{4 Q^2}{r^2} {\cal H}_P
\end{array} \right). \nonumber
\end{eqnarray}
As we will show in the next section in more detail, stability follows
from the fact that the spatial part of the operator $\Dtens$ and
the matrix $\Stens$ are manifestly positive.

\subsubsection{Pulsation equations ($\ell=1$)}

In the sector $\ell=1$, $\lambda$ vanishes and
equation (\ref{Eq:Odd1}) can be integrated to yield
\begin{displaymath}
r^4\asttil d\left( \frac{h}{r^2} \right) = 4Q\, \nu + 6\, J,
\end{displaymath}
where $J$ is an integration constant. 
Using this into equation (\ref{Eq:Odd2}) gives
\begin{equation}
{\cal H}_P \codtil\left[ {\cal H}_P^{-1} d\nu \right] 
 + \left( \frac{2}{r^2} + 4{\cal H}_P\frac{Q^2}{r^4} \right)\nu = 
 - 6{\cal H}_P \frac{Q\, J}{r^4}\, .
\label{Eq:Pulsl1}
\end{equation}
The general solution to this equation is the sum of a particular
solution $\nu_1$ and the general homogeneous solution.
The particular solution, which is proportional to $J$, can
be chosen to be a function of $r$ only. Therefore, $\nu_1$
represents a stationary axial excitation of the background
configuration. For ${\cal H}_P = 1$ an explicit solution 
is given by $\nu_1 = J/r$ which corresponds to the Kerr-Newman 
solution to linear order in the rotation parameter $J/M$ \cite{S-Diss}. 
For a NED, we have not found an explicit expression for $\nu_1$,
but we anticipate that $\nu_1$ may give rise to slowly rotating
generalizations of the solutions found in \cite{ABG-E1, ABG-E2, ABG-E3}.
The homogeneous solutions represent electromagnetic radiation
which propagates in a stable way since the potential on the
left-hand side of equation (\ref{Eq:Pulsl1}) is positive
(see also the next section).

\subsection{Even-parity sector}

Gravitational perturbations are more complicated in the even-parity sector
than in the odd-parity one.  The metric is now parametrized
by a symmetric $2$ tensor, $H_{ab}$, a one-form $Q_a$ and two
scalar fields $K$ and $G$ in the following way:
\begin{eqnarray}
\delta g_{ab} &=& H_{ab}\, Y, \nonumber\\
\delta g_{aB} &=& Q_a\, \nabhat_B Y, \nonumber\\
\delta g_{AB} &=& K\, g_{AB} Y 
 + G r^2 \left( \nabhat_A\nabhat_B - \frac{1}{2}\,\ghat_{AB}\laphat \right)Y,
\nonumber
\end{eqnarray}
where $G$ is not present for $\ell=1$ since 
$\left( \nabhat_A\nabhat_B - \frac{1}{2}\,\ghat_{AB}\laphat \right)Y = 0$
in this case.
For $\ell\geq 2$ one can introduce the following gauge invariants
\cite{GS}:
\begin{displaymath}
H_{ab}^{(inv)} = H_{ab} - 2\nabtil_{(a} p_{b)}\; ,\qquad 
K^{(inv)} = K - 2\frac{p^a\nabtil_a r}{r} + \frac{1}{2}\ell(\ell+1) G,
\end{displaymath}
where $p_a = Q_a - (r^2\nabtil_a G)/2$.
With respect to an infinitesimal coordinate transformation
parametrized by the vector field 
$X = \xi^a Y \partial_a + f\nabhat^A Y\partial_A$ we have
\begin{equation}
p_a \mapsto p_a + \xi_a\; ,\qquad
G \mapsto G + 2f.
\label{Eq:CoordTransEven}
\end{equation}
Since $G$ and $p_a$ can both be set to zero by an appropriate
coordinate transformation, we can assume that these quantities
vanish when deriving the pulsation equations below.
In this gauge, $H_{ab}^{(inv)} = H_{ab}$, $K^{(inv)} = K$ and
$Q_a$ and $G$ vanish. Since the perturbation equations are
gauge invariant, it is sufficient to replace $H_{ab}$ by
$H_{ab}^{(inv)}$ and $K$ by $K^{(inv)}$ in the final expressions
and so obtain the equations in an arbitrary gauge.

The electromagnetic potential is parametrized according to
\begin{displaymath}
\delta A_\mu dx^\mu = \alpha\, Y + \mu\, d Y,
\end{displaymath}
with $\alpha = \alpha_b dx^b$ a one-form and 
$\mu$ a scalar on $\tilde{M}$.
The corresponding electromagnetic field is
\begin{displaymath}
\delta F_{\mu\nu} = \partial_\mu \delta A_\nu - \partial_\nu \delta A_\mu
 = \left[ d\alpha\, Y + (d\mu - \alpha) \wedge d Y \right]_{\mu\nu},
\end{displaymath}
from which it is clear that the amplitude 
$\hat{\alpha} = \alpha - d\mu$ is $U(1)$-invariant.
Using the background expression (\ref{Eq:Ptensback}) and 
$\pounds_X\Ftens = d i_X\Ftens = d (r^{-2}{\cal H}_P Q \etatil_{ab} X^a dx^b)$
we find that with respect to an infinitesimal change of coordinates,
\begin{displaymath}
\hat{\alpha} \mapsto \hat{\alpha} + {\cal H}_P\frac{Q}{r^2}\, \asttil(\xi_a dx^a).
\end{displaymath}
In view of the transformations (\ref{Eq:CoordTransEven})
we see that we can construct a coordinate invariant one-form
which reduces to $\hat{\alpha}$ if $p_a = 0$.

\subsubsection{Electromagnetic equations}

Having discussed the gauge issues, we now proceed to the
derivation of the perturbation equations. First, a small
calculation using (\ref{Eq:LinPtens}) yields
\begin{displaymath}
\delta P_{\mu\nu} = \left[ \asttil\pi\, Y + \gamma \wedge d Y \right]_{\mu\nu}\, ,
\end{displaymath}
with
\begin{eqnarray}
\pi &=& -\frac{{\cal H}_P^{-1}}{\kappa}\, \asttil d\hat{\alpha} 
 + \frac{\kappa-1}{2\kappa}\frac{Q}{r^2}\, H ,
\label{Eq:pi}\\
\gamma &=& -{\cal H}_P^{-1} \hat{\alpha}, 
\label{Eq:gamma}
\end{eqnarray}
where $\kappa$ is defined in (\ref{Eq:KDef}) and $H = H^a_{\; a}$.
In terms of the scalar $\pi$ and the one-form $\gamma$
we can write the electromagnetic equations as
\begin{eqnarray}
\asttil d\left[ -r^2\pi + \frac{Q}{2}(H - 2K) \right] - \ell(\ell+1)\gamma
&=& 0, \label{Eq:EM1}\\
\codtil\gamma &=& 0. \label{Eq:EM2}
\end{eqnarray}
Equation (\ref{Eq:EM2}) is the integrability condition for 
equation (\ref{Eq:EM1}) and can be used in order to introduce a 
scalar potential $\phi$ according to
\begin{equation}
\gamma = \asttil d\phi.
\label{Eq:Defphi}
\end{equation}
Equation (\ref{Eq:EM1}) can then be integrated to obtain
the algebraic relation\footnote{The integration constant is absorbed into the definition of $\phi$.}
\begin{displaymath}
r^2\pi = -\ell(\ell+1)\phi + \frac{Q}{2}(H - 2K).
\end{displaymath}
Using the equations (\ref{Eq:pi}) and (\ref{Eq:gamma})
one then obtains the following pulsation equations for the
electromagnetic field
\begin{equation}
{\cal H}_P^{-1}\codtil\left( {\cal H}_P d\phi \right) 
 + \kappa\frac{\ell(\ell+1)}{r^2}\,\phi
 + \frac{Q}{2r^2}\left( 2\kappa K - H \right) = 0.
\label{Eq:C1}
\end{equation}
If we turn off the gravitational amplitudes (i.e. if
$H = K = 0$), we see that the sign of the potential
depends on the function $\kappa$ defined in (\ref{Eq:KDef}).
In particular, the equations are stable as long as $\kappa \geq 0$.
In the next section we will show that this is also a necessary condition
for stability, i.e. we will show that if $\kappa < 0$ in
an interval, the equations admit exponentially growing modes
for sufficiently large $\ell$.

\subsubsection{Gravitational equations}

Since the electromagnetic field couples back to 
the gravitational field, we have to consider the linearized 
Einstein equations (\ref{Eq:LinEinstein}) as well. The
harmonic decomposition of the linearized Einstein tensor
can be found in \cite{S-Diss}. The result is quite complicated,
but the structure of the equations becomes somewhat more
transparent if one splits the tensor $H_{ab}$ into its
trace and traceless part and then introduces the one-form
\begin{displaymath}
C = H^{TF}_{ab}(\nabtil^a r) d x^b,
\end{displaymath}
where $H^{TF}_{ab} = H_{ab} - \frac{1}{2}\gtil_{ab} H$ 
denotes the traceless part of $H_{ab}$.
The relevant components of the linearized Einstein tensor 
may then be expressed in the following way:
\begin{eqnarray}
(\delta G)^{TF}_{AB} &=& S \left( \nabhat_A\nabhat_B - \frac{1}{2}\ghat_{AB}\laphat \right) Y,
\nonumber\\
\gtil^{ab}\delta G_{ab} &=& \left( T + G^{ab}H^{TF}_{ab} \right) Y,
\nonumber\\
\delta G_{Ab}\,d x^b &=& \left( U + U_{BG} \right)  \frac{1}{2}\nabhat_A Y, 
\nonumber\\
(\delta G)^{TF}_{ab} r^{|a} d x^b &=& \left( V - \frac{1}{2} G^{ab}H^{TF}_{ab}\,d r + G_{ab}(C^a - \frac{r}{2} \nabtil^a K)d x^b \right) Y, \nonumber
\end{eqnarray}
where
\begin{displaymath}
U_{BG} = \frac{r}{2N} \left( G^{ab} d r + (\etatil^{ac} G^b_{\; c}) \asttil d r \right) H^{TF}_{ab}\; ,
\end{displaymath}
and
\begin{eqnarray}
S &=& -\frac{1}{2} H, \label{Eq:LinG1}\\
T &=& \frac{2}{r} \codtil C - \frac{2}{r^2} \gtiltens(C, d r) 
   + \laptil K + \frac{4}{r}\gtiltens(d K, d r)
   - \frac{\lambda}{r^2} K - \frac{\lambda + 4}{2r^2} H, 
\label{Eq:LinG2}\\
U &=& -\frac{1}{N} \left[ (\codtil C) d r 
   + (\asttil d C) \asttil d r \right]
   -d\left(K + \frac{1}{2} H \right) + H\,\frac{d r}{r}, 
\label{Eq:LinG3}\\
V &=& (\codtil C) \frac{d r}{r} + \frac{1}{r} d\gtiltens( C, d r) 
   + \frac{\ell(\ell+1)}{2r^2} C 
\nonumber\\
  &+& \frac{1}{2}\laptil K\,d r - d \gtiltens( d K, d r) 
   + \left(\laptil r - \frac{N+1}{2r} \right)d K + \frac{N}{2r} d H. 
\label{Eq:LinG4}
\end{eqnarray}
Here, $N = (\nabtil^a r)(\nabtil_a r)$ agrees with the function
$N$ defined in the previous section, see equation (\ref{Eq:SchwCoord}).
Note that equation (\ref{Eq:LinG1}) is void when $\ell=1$ since
then $\nabhat_A\nabhat_B Y = -3\ghat_{AB} Y$.

In our case $G_{ab} = 2\gtil_{ab} {\cal H}$ is proportional
to $\gtil_{ab}$, so we have $G^{ab} H^{TF}_{ab} = 0$ and $U_{BG} = 0$.
Next, it is not difficult to see from the right-hand side of equation 
(\ref{Eq:LinEinstein}) that $\delta G_{AB}$ is proportional
to $g_{AB}$ so that $S = 0$. Then, it follows from equation
(\ref{Eq:LinG1}) that the trace of $H_{ab}$ vanishes, $H = 0$.
(When $\ell=1$, $H$ can be set to zero by an appropriate coordinate transformation).
The remaining components of the right-hand side of (\ref{Eq:LinEinstein})
yield the expressions
\begin{eqnarray}
T &=& -4 {\cal H}_P\frac{Q}{r^2}\,\pi, \label{Eq:LinT2}\\
U &=& 4 {\cal H}_P\frac{Q}{r^2}\,d\phi, \label{Eq:LinT3}\\
V &=& {\cal H}\, rd K, \label{Eq:LinT4}
\end{eqnarray}
where equation (\ref{Eq:Defphi}) has been used in
the derivation for $U$. Following the vacuum case \cite{S-Diss, ST}
we first look at the component $\gtiltens(U,\asttil dr)$:
Using equations (\ref{Eq:LinG3}) and (\ref{Eq:LinT3}), we
find
\begin{displaymath}
\asttil d [ C - rd K ] = 
 -4\asttil d\left[ {\cal H}_P\phi\,\frac{Q}{r^2} dr \right],
\end{displaymath}
where we have used the identity $\gtiltens(\alpha,\asttil dr) =
-\asttil(\alpha\wedge dr)$ for any one-form $\alpha$.
This motivates the introduction of a one-form $Z$ defined by
\begin{displaymath}
Z = C - r dK + 4 {\cal H}_P\phi\,\frac{Q}{r^2} dr.
\end{displaymath}
Since $d Z = 0$ we can introduce a potential $\zeta$
with $Z = d\zeta$.
Next, we look at the combination $2V - T dr$. Using equations
(\ref{Eq:LinG2},\ref{Eq:LinG4}) and (\ref{Eq:LinT2},\ref{Eq:LinT4}),
we find
\begin{eqnarray}
&& d\left[ 2r\gtiltens\left( Z - 4{\cal H}_P\phi\, \frac{Q}{r^2}d r,d r \right) \right] + \ell(\ell+1)Z \nonumber\\
&+& r\left( \lambda + 3(1-N) + 2r^2{\cal H} \right) d K
   + \left(\lambda + 4{\cal H}_P\frac{Q^2}{r^2} \right) K d r = 0.
\label{Eq:C2}
\end{eqnarray}
By virtue of the background equations we have
\begin{eqnarray}
\partial_r\left[ r\left(\lambda + 3(1-N) + 2r^2{\cal H}\right) \right]
 &=& \lambda + 6 \partial_r m + 6r^2{\cal H} 
  + 2r^3 {\cal H}_P \partial_r P \nonumber\\
 &=& \lambda + 4{\cal H}_P\frac{Q^2}{r^2}\; , \nonumber
\end{eqnarray}
and since $Z = d\zeta$, we can integrate equation (\ref{Eq:C2}) to 
obtain\footnote{The integration constant is absorbed into the definition of $\zeta$.}
\begin{equation}
2r\gtiltens(d\zeta,d r) + \ell(\ell+1)\zeta 
 + \left(r\lambda + 6m + 2r^3{\cal H}\right) K 
 - 8N {\cal H}_P\frac{Q}{r}\, \phi = 0.
\label{Eq:C3}
\end{equation}
Finally, we compute the combination $r T + \gtiltens(U,d r)$
which yields, using the equations (\ref{Eq:LinG2},\ref{Eq:LinG3}) and 
(\ref{Eq:LinT2},\ref{Eq:LinT3}),
\begin{equation}
\codtil Z - \frac{2}{r}\, \gtiltens(Z,d r) 
  - \left( \frac{\lambda}{r} + 4{\cal H}_P\frac{Q^2}{r^3} \right) K
 + 4\left[ -\codtil\left( {\cal H}_P\frac{Q}{r^2}d r \right) 
 + 2N{\cal H}_P\frac{Q}{r^3} - \frac{\ell(\ell+1)}{r^2}\frac{Q}{r}\,{\cal H}_P \right]\phi = 0.
\label{Eq:C4}
\end{equation}
This equation may be used in order to express the scalar $K$
in terms of the other perturbation amplitudes. 

To summarize, we
have the perturbation equations (\ref{Eq:C1}) for the
electromagnetic perturbation potential $\phi$ and equation
(\ref{Eq:C3}) for the gravitational perturbation potential $\zeta$
where $H=0$ and $K$ can be eliminated using equation (\ref{Eq:C4}).
The next task is to bring these equations in form of a wave-like
equation with symmetric potential which is more suitable for a
discussion of stability.

\subsubsection{Gravito-electromagnetic equations}

The pulsation equations can be brought into the form of
a wave-like equation with symmetric potential after the
following transformations:
\begin{equation}
\Psi = \sqrt{\lambda}\,\frac{ \zeta }{a + \lambda}\; ,\qquad
\Phi = \sqrt{4{\cal H}_P} \left( \phi - \frac{Q}{r}\,\frac{ \zeta }{a + \lambda} \right),
\label{Eq:NewPot}
\end{equation}
where $a = 3r\laptil r - 4r^2{\cal H} = 6m/r + 2r^2 {\cal H}$.
For the moment, we restrict ourselves to the case $\ell\geq 2$.
Note that $a$ is positive outside the horizon of a regular black hole
(see below).

In terms of the new potentials defined in (\ref{Eq:NewPot}), 
a long calculation yields
the final pulsation equations
\begin{equation}
-\laptil\left( \begin{array}{c} \Psi \\ \Phi \end{array} \right)
+ \left( \begin{array}{cc}
  V_{11} & -\frac{ \sqrt{4\lambda {\cal H}_P}\, Q}{r^3}\, W \\
  -\frac{ \sqrt{4\lambda {\cal H}_P}\, Q}{r^3}\, W & V_{22} 
  \end{array} \right)
\left( \begin{array}{c} \Psi \\ \Phi \end{array} \right) = 0,
\label{Eq:PulsEven}
\end{equation}
where
\begin{eqnarray}
V_{11} &=& \frac{1}{r^2(a + \lambda)}\left[ \ell(\ell+1)\lambda -2N\lambda + a\, r\laptil r \right] + \frac{2N\lambda\, b}{r^2(a + \lambda)^2}\, , \nonumber\\
V_{22} &=& \kappa\frac{\ell(\ell+1)}{r^2}
 + \frac{ 4{\cal H}_P\, Q^2}{r^4(a + \lambda)}\left( \lambda + 1 - N - 2r^2 {\cal H} + 4N\kappa \right)
 + {\cal H}_P^{-1/2}\laptil {\cal H}_P^{1/2}
 + \frac{8N {\cal H}_P\, Q^2\, b}{r^4(a + \lambda)^2}\, , \nonumber\\
W &=& \frac{1}{a + \lambda}\left( \lambda + 1 - N - 2r^2 {\cal H} + 2N\kappa \right) + \frac{2N\, b}{(a + \lambda)^2}\, ,
\nonumber
\end{eqnarray}
and where we have defined
\begin{displaymath}
b = \lambda + 4{\cal H}_P \frac{Q^2}{r^2}\, .
\end{displaymath}
Here, we have also used the background expressions
\begin{eqnarray}
r\laptil r = r\partial_r N &=& 1 - N + 2r^2 {\cal H}, \nonumber\\
r{\cal H}_P^{-1}\codtil({\cal H}_Pd r) &=& -1 - N - 2r^2 {\cal H} + 2N\kappa.
\nonumber
\end{eqnarray}
The equations (\ref{Eq:PulsEven}) generalize the pulsation equations
obtained by Moncrief \cite{Moncrief} to NED
coupled to gravity. (For ${\cal H} = P$ the equations above reduce
to the one obtained in \cite{Moncrief}.)

\subsubsection{Pulsation equations ($\ell\geq 2$)}

We now proceed to show under which conditions the potential in
the equation (\ref{Eq:PulsEven}) is positive semi-definite.

We first show that the function $a = 6m/r + 2r^2{\cal H}$ is positive
outside an event horizon:
Let $r=r_h$ denote the radius of the event horizon. Then,
$m(r_h) = r_h/2$ and  $0 \leq r_h\partial_r N(r_h) = 2r_h^2 {\cal H}(r=r_h) + 1$.
Therefore, $a(r=r_h) \geq 2$. Next, using the field equations 
$\partial_r m = -r^2 {\cal H}$, we have
$\partial_r (ra) = 4{\cal H}_P Q^2/r^2 > 0$. This shows that $a$ cannot
become negative or zero for $r > r_h$.
In order to prove that the potential is positive, we first absorb
a term $r^{-1}\laptil r$ appearing in $V_{11}$ and the term 
${\cal H}_P^{-1/2}\laptil {\cal H}_P^{1/2}$ appearing in $V_{22}$
into the differential operator. We may then rewrite the pulsation
equations into the form
\begin{equation}
\left( \Dtens + \Stens \right)
\left( \begin{array}{c} \Psi \\ \Phi \end{array} \right) = 0, 
\label{Eq:PulsEven2}
\end{equation}
where
\begin{eqnarray}
\Dtens &=& \left( \begin{array}{cc} 
 -\frac{1}{r}\nabtil^a r^2 \nabtil_a \frac{1}{r} & 0 \\
 0 & -{\cal H}_P^{-1/2} \nabtil^a {\cal H}_P \nabtil_a {\cal H}_P^{-1/2} 
\end{array} \right), 
\nonumber\\
\Stens &=& \left( \begin{array}{cc}
 \frac{\lambda}{r^2(a + \lambda)}\left[ c_1 + \frac{2Nb}{a+\lambda} \right] &
 -\frac{\sqrt{4\lambda {\cal H}_P}\, Q}{r^3(a+\lambda)}\left[ w + \frac{2Nb}{a+\lambda} \right]  \\
 -\frac{\sqrt{4\lambda {\cal H}_P}\, Q}{r^3(a+\lambda)}\left[ w + \frac{2Nb}{a+\lambda} \right] &
\kappa\frac{\ell(\ell+1)}{r^2} + \frac{4{\cal H}_P Q^2}{r^4(a + \lambda)}
\left[ c_2 + \frac{2Nb}{a+\lambda} \right]
\end{array} \right), \nonumber
\end{eqnarray}
with
\begin{eqnarray}
c_1 &=& \lambda + 1 - N - 2r^2{\cal H}, \nonumber\\
c_2 &=& c_1 + 4N\kappa, \nonumber\\
w &=& c_1 + 2N\kappa. \nonumber
\end{eqnarray}
Assume that ${\cal H} \leq 0$ and $\kappa\geq 0$ for all $r > r_h$.
It then follows that $a$ ,$b$ and $c_1$ are positive since $N \leq 1$.
Therefore, the diagonal elements of $\Stens$ are positive. 
In order to show that $\Stens$ is positive definite, it remains to check 
the sign of its determinant. A small calculation yields
\begin{displaymath}
\frac{r^4(a+\lambda)^2}{\lambda}\,\det\Stens =
\kappa\ell(\ell+1)\left[ (a+\lambda)c_1 + 2\lambda N \right]
 + 8N\kappa{\cal H}_P\frac{Q^2}{r^2}\left[ \ell(\ell+1) - 2N\kappa \right].
\end{displaymath}
This is manifestly positive if
\begin{equation}
0 < 2N\kappa \leq \ell(\ell+1)
\label{Eq:StabCond1}
\end{equation}
(if $\kappa=0$ $\Stens$ is only positive semi-definite.)
In particular, this condition is satisfied in linear electrodynamics
since then $\kappa = 1$. Provided that the weak field limit holds
and that $\kappa$ is regular and positive near the event horizon,
we have $N\kappa\rightarrow 1$ as $r\rightarrow\infty$
and $N\kappa\rightarrow 0$ as $r\rightarrow r_h$, so in those limits
the condition (\ref{Eq:StabCond1}) is satisfied.

\subsubsection{Pulsation equations ($\ell=1$)}

For $\ell=1$ it does not make sense to rescale $\zeta$ by
$\sqrt{\lambda}$ since $\lambda = 0$ in this case. Considering
$\Psi = \zeta/a$ instead of the $\Psi$ defined in (\ref{Eq:NewPot}),
we see that the equation for $\Phi$ decouples when $\ell=1$.
Repeating the arguments above, it is easy to check that the 
corresponding potential is positive if $\kappa > 0$.
The equation for $\Psi$ can be rewritten as
\begin{equation}
\frac{1}{r}\nabtil^a \left[ r^2\nabtil_a\left( \frac{\Psi}{r} \right) \right]
=  -\frac{\sqrt{4{\cal H}_P}\, Q}{r^3\, a}\left[ w + \frac{2Nb}{a} \right]\Phi.
\label{Eq:PulsEven3}
\end{equation}
On the other hand, the perturbation amplitudes are not necessarily
gauge invariant when $\ell=1$. Indeed, as shown in \cite{S-Diss},
the potentials can be seen to transform according to
\begin{displaymath}
\zeta \mapsto \zeta + ra\, f, \qquad
\phi \mapsto \phi + Q\, f
\end{displaymath}
with respect to an infinitesimal coordinate transformation
parametrized by the vector field 
$X = -r^2\nabtil^a f Y\partial_a + f \nabhat^A Y\partial_A$
where $f$ satisfies $\nabtil^a(r^2\nabtil_a f) = 0$
(This makes sure that we remain in a gauge with $H=0$.)
Therefore, $\Psi \mapsto \Psi + r\, f$ while the electromagnetic
potential $\Phi$ is gauge invariant. Having solved for $\Phi$,
equation (\ref{Eq:PulsEven3}) can be used in order to obtain
the gauge-invariant part of $\Psi$.

\section{Stability analysis}
\label{Sect-4}

In the previous section, we have shown that linear fluctuations
around a spherically symmetric and purely electric solution 
are governed by a wave-like equation with symmetric potential
of the form
\begin{equation}
\left( -\Ptens\nabtil^a \Ptens^{-2} \nabtil_a\Ptens + \Stens \right) u = 0,
\label{Eq:WaveStab}
\end{equation}
where $\nabtil$ denotes the covariant derivative with respect 
to the metric $\gtiltens$ and where $\Ptens$ is a positive definite 
symmetric matrix and $\Stens$ is a symmetric matrix. 
The vector-valued function $u$ on $\tilde{M}$ is a gauge-invariant 
combination of the perturbed metric and electromagnetic fields.
We will show below that linear stability follows if the potential
$\Stens$ is positive definite. In the previous section,
we have shown that this is the case if the following conditions
hold outside the event horizon:
\begin{eqnarray}
&& {\cal H} < 0, \qquad
   {\cal H}_P > 0, 
\label{Eq:WEC}\\
&& 0 < N\kappa \leq 3,
\label{Eq:StabCond}
\end{eqnarray}
where $\kappa = 1 + 2{\cal H}_P^{-1} {\cal H}_{PP} P$.
The first two conditions (\ref{Eq:WEC}) imply that the 
weak energy condition is satisfied outside the horizon,
while condition ({\ref{Eq:StabCond}) implies that the function
$F(P)$ is well defined outside the horizon, see (\ref{Eq:FPEquiv}).
We have checked that these conditions are verified for
the Hamiltonian (\ref{Eq:ABG-E1Ham}) used in \cite{ABG-E1}, 
for several values of the parameter $s$.
However, it does not seem to be difficult to modify
the function ${\cal H}(P)$ slightly and to violate,
for example, the condition $\kappa > 0$ in a small
interval outside the event horizon. As we will
show below, the background solution is unstable if
$\kappa < 0$ at some point outside the event horizon.

One way of analyzing the behavior of the solutions to
equation (\ref{Eq:WaveStab}) is by introducing
Schwarzschild coordinates $(t,r)$, see equation (\ref{Eq:SchwCoord}),
in which case (\ref{Eq:WaveStab}) can be written as
\begin{equation}
\left(\partial_t^{\; 2} - \Ptens\partial_{r_*} \Ptens^{-2} \partial_{r_*} \Ptens
 + N\Stens \right) u = 0,
\label{Eq:WaveStab1}
\end{equation}
where $\partial_{r_*} \equiv N\partial_r$ denotes the derivative with
respect to the tortoise coordinate.
(Notice that $\Ptens$ does not depend on $t$.) 
One then considers perturbations
which are smooth and vanish at the boundary points
$r=r_h$ (${r_*}=-\infty$) and $r=\infty$ (${r_*}=\infty$).
The spatial part of the wave operator in (\ref{Eq:WaveStab1})
is symmetric with respect to the $L^2$ product
\begin{displaymath}
(u,v) = \int_{-\infty}^{\infty} u^\dagger v\, d{r_*}
\end{displaymath}
for such perturbations. Since that operator is also real,
self-adjoint extensions exist (in this case the operator is
self-adjoint on some Sobolev space), and the linear 
stability can be analyzed along the lines of \cite{Wald, VG}
using the tools from spectral theory. 
As a first step one can look at solutions of 
(\ref{Eq:WaveStab1}) with time dependency of the form 
$u \sim e^{i\omega t}$, in which case
\begin{equation}
\omega^2 u = \left( -\Ptens\partial_{r_*} \Ptens^{-2} \partial_{r_*} \Ptens 
 + N\Stens \right) u.
\label{Eq:SturmLiouville}
\end{equation}
If the operator on the right-hand side of (\ref{Eq:SturmLiouville})
is positive, which is the case if $\Stens$ is positive definite, 
it follows that $\omega$ is real and positive and there are no 
unstable modes (if $\omega=0$ there might be modes with linear growth in $t$).
On the other hand, if $\Stens$ is negative the operator
is not necessarily positive. If one can find an eigenfunction
with negative eigenvalue, $\omega$ is complex and one has
a solution that grows exponentially in time.
While it is difficult to solve the eigenvalue problem explicitly
in general, there are simple analytical methods to show the
existence of at least one unstable mode \cite{VG}.
In fact, if one can find a function $u$ with $(u,u) < \infty$
and $(u,Hu) < 0$, this establishes the existence of at least one
negative eigenvalue and the background solution is linearly
unstable.

However, the analysis of the eigenvalue problem (\ref{Eq:SturmLiouville})
only gives information on the point spectrum
(or on quasi-normal frequencies, but the corresponding modes $u$
do not satisfy $(u,u) < \infty$).
Generalizing the results obtained in \cite{Wald} on should
be able to show the following stronger stability result:
If the potential $N\Stens$ is strictly positive for all $r > r_h$
and if $u(t=0,.)$, $\partial_t u(t=0,.)$ are smooth and have
compact support, there is a bound $C < \infty$ such that
\begin{displaymath}
| u(t,r_*) | \leq C
\end{displaymath}
for all $t \geq 0$ and $-\infty < r_* < \infty$.
(If $N\Stens$ is nonnegative but fails to be strictly positive
outside the event horizon, the perturbations might exhibit
linear growth in time.)

We now use the above arguments to show that if $\kappa < 0$,
the solution is unstable:
Suppose that $\kappa_1 \equiv \kappa(r_1) < 0$ for some $r_1 > r_h$.
Assuming that $\kappa$ is continuous, there is a $\delta > 0$
small enough such that $\kappa < \kappa_1/2$ on the interval 
$(r_1 - \delta, r_1 + \delta)$ and $\kappa < 0$ on the interval
$(r_1 - 2\delta, r_1 + 2\delta)$.
Consider the test function
\begin{displaymath}
u = (\Psi,\Phi) = \left( \sqrt{4\lambda^{-1}{\cal H}_P}\, Q f(r), rf(r) \right), 
\end{displaymath}
for equation (\ref{Eq:PulsEven2})
where the function $f(r)$ is smooth, vanishes identically outside
the interval $(r_1-2\delta,r_1+2\delta)$ and is identically one for 
$r\in (r_1-\delta, r_1+\delta)$. 
It is clear that $(u,u) < \infty$, and
$u^\dagger\Stens u = \kappa\,\ell(\ell+1) f^2$.
Furthermore, a small calculation gives
\begin{displaymath}
(u,Hu) \leq 4\delta\left( 1 + \frac{4}{\lambda} \right) K^2 
 - \delta |\kappa_1| \ell(\ell+1),
\end{displaymath}
where $K$ is a constant which is independent on $\ell$ but
large enough, such that
\begin{displaymath}
\Big| Q r\partial_{r_*}\left( {\cal H}_P^{1/2} r^{-1} f\right) \Big| \leq K, \qquad
\Big| {\cal H}_P^{1/2} \partial_{r_*}\left( {\cal H}_P^{-1/2} r f\right) \Big| \leq K.
\end{displaymath}
It is clear that $(u,Hu) < 0$ for large enough angular momentum
number $\ell$. Therefore, the background configuration is
linearly unstable if $\kappa$ is negative somewhere outside
the event horizon.

\section{Conclusion}
\label{Sect-5}

We have derived the pulsation equations governing linear
fluctuations on a spherically symmetric and electrically 
charged black hole solution in NED coupled to gravity. 
Under a mild technical assumption (namely that ${\cal H}_P$ 
vanishes nowhere outside the event horizon) the pulsation 
equations in the odd-parity sector cannot exhibit modes that
grow exponentially in time outside the event horizon.
However, in the even-parity sector, the stability properties
depend strongly on the Hamiltonian ${\cal H}(P)$, and one
does not necessarily have the equivalence between the spectra
in the odd- and even-parity sectors via a supersymmetric
transformation like for linear perturbations in Einstein-Maxwell
theory \cite{C-Book}. As a consequence, the quasi-normal
frequencies might be different in the odd- and even-parity sectors
if the electromagnetic Lagrangian is not linear.

In terms of the dimensionless variable 
$x = \sqrt{-2Q^2\, P} = Q^2/r^2$, sufficient conditions
for linear stability are
\begin{eqnarray}
{\cal H} &<& 0, \label{Eq:Cond1-E}\\
{\cal H}_x &<& 0,  \label{Eq:Cond2-E}\\
{\cal H}_{xx} &<& 0, \label{Eq:Cond3-E}\\
3{\cal H}_x &\leq& xN{\cal H}_{xx}\; , \label{Eq:Cond4-E}
\end{eqnarray}
for all $0 < x < x_h$ (i.e. $r > r_h$), where $x_h = Q^2/r_h^2$ 
corresponds to the value of $x$ at the event horizon.
While we have shown that a solution is unstable if
${\cal H}_{xx} > 0$ for some $x < x_h$ (the case
${\cal H}_{xx} \geq 0$ corresponding to a marginal case
where growth which is linear in time might appear)
we stress that the conditions 
(\ref{Eq:Cond1-E},\ref{Eq:Cond2-E},\ref{Eq:Cond4-E}) 
are only sufficient. For example, there might exist stable 
black holes that satisfy ${\cal H}_{xx} < 0$
but violate the condition (\ref{Eq:Cond4-E}).

The linear stability of the purely magnetic solutions 
considered in \cite{ABG-M1, Br} can be discussed on the same
lines as above. In fact, using the $FP$ duality transformation
(\ref{Eq:FPDuality}), one can easily translate the conditions
(\ref{Eq:Cond1-E},\ref{Eq:Cond2-E},\ref{Eq:Cond3-E},\ref{Eq:Cond4-E}) 
to obtain the sufficient conditions
\begin{eqnarray}
{\cal L} &>& 0, \label{Eq:Cond1-M}\\
{\cal L}_y &>& 0,  \label{Eq:Cond2-M}\\
{\cal L}_{yy} &>& 0, \label{Eq:Cond3-M}\\
3{\cal L}_y &\geq& yN{\cal L}_{yy}\; , \label{Eq:Cond4-M}
\end{eqnarray}
where $y = \sqrt{2g^2F}$ and $g$ is the magnetic charge.

One can show that the conditions 
(\ref{Eq:Cond1-M},\ref{Eq:Cond2-M},\ref{Eq:Cond3-M},\ref{Eq:Cond4-M})
are satisfied for the Bardeen black holes of \cite{ABG-M1} for
any value of the charge to mass ratio (such that one has a black hole).
A proof of this statement is given in the appendix.
We have also checked that the conditions 
(\ref{Eq:Cond1-E},\ref{Eq:Cond2-E},\ref{Eq:Cond3-E},\ref{Eq:Cond4-E}) 
hold for the black hole solutions found in \cite{ABG-E1,ABG-E2,ABG-E3} 
for different values of the charge to mass ratio. 
This establishes the linear stability of these solutions
with respect to linear fluctuations in the domain of outer
communication.
An interesting question is what happens to perturbations
inside the black hole. As expected from the mass inflation
scenario \cite{PI} fluctuations that are falling into the
black hole are likely to destroy the Cauchy horizon and 
to introduce a singularity inside the hole.
These questions will be analyzed further in some future work
where we also plan to discuss the stability of particle-like
solutions \cite{ABMS}.

\section*{Acknowledgments}
The authors wish to thank Eloy Ay\'on-Beato and
Carlos Lousto for many useful discussions. 
This work was supported by the
Mexican Council of Science and Technology, CONACyT, 
and the Swiss National Science Foundation.

\appendix
\section{Linear stability of the Bardeen black holes}

The Bardeen black holes are obtained in the $F$ framework 
from the Lagrangian \cite{ABG-M1}
\begin{displaymath}
{\cal L}(y) = \frac{3}{2sg^2}\left( \frac{y}{1+y} \right)^{5/2},
\end{displaymath}
where $y = \sqrt{2g^2F}$ and
$s = |g|/2m$ is half the monopole charge to mass ratio.
Since
\begin{eqnarray}
{\cal L}_y &=& \frac{15}{4sg^2}\frac{ y^{3/2} }{(1+y)^{7/2}}\; , 
\nonumber\\
{\cal L}_{yy} &=& \frac{15}{8sg^2} \frac{ y^{1/2} }{(1+y)^{9/2}} (3-4y), 
\nonumber\\
f(y) \equiv y {\cal L}_y^{-1} {\cal L}_{yy} &=& \frac{3-4y}{2(1+y)}\; ,
\nonumber
\end{eqnarray}
it is clear that ${\cal L}_y > 0$ and so it is sufficient to
check the condition 
\begin{equation}
3\geq Nf(y) > 0
\label{Eq:InEq}
\end{equation}
for $0\leq y \leq y_h$ where $y_h$ is the value of $y$ at the
event horizon.

Since $f(y)$ is monotonously decreasing with $f(0) = 3/2$ the
first part of the inequality (\ref{Eq:InEq}) is verified.
Next, consider the component
\begin{displaymath}
N = -g_{tt} = 1 - \frac{1}{s} \frac{y^{1/2}}{(1+y)^{3/2}}
\end{displaymath}
of the metric which determines the location of the event horizon.
$N$ has a single minimum at $y=1/2$, so $y_h \leq 1/2$. 
Therefore, the second part of the inequality (\ref{Eq:InEq}) 
is also verified for all $0 \leq y \leq y_h$ and we conclude
that the Bardeen black holes are stable.




\begin{thebibliography}{10}



\bibitem{ABG-E1}
E. Ay\'on-Beato and A. Garc\'{\i}a,
Phys. Rev. Lett. {\bf 80}, 5056 (1998).

\bibitem{ABG-E2}
E. Ay\'on-Beato and A. Garc\'{\i}a,
Phys. Lett. B {\bf 464}, 25 (1999).

\bibitem{ABG-E3}
E. Ay\'on-Beato and A. Garc\'{\i}a,
Gen. Rel. Grav. {\bf 31}, 629 (1999).

\bibitem{ABG-M1}
E. Ay\'on-Beato and A. Garc\'{\i}a,
Phys. Lett. B {\bf 493}, 149 (2000).

\bibitem{Br}
K.A. Bronnikov,
Phys. Rev. D {\bf 63}, 044005 (2001).

\bibitem{BH}
A. Burinskii and S.R. Hildebrandt,
Phys. Rev. D {\bf 65}, 104017 (2002).

\bibitem{EH}
E. Elizalde and S.R. Hildebrandt,
Phys. Rev. D {\bf 65}, 124024 (2002).

\bibitem{HE}
S.W. Hawking and G.F.R. Ellis,
{\em The large scale structure of space-time},
Cambridge University Press (1973)
and references therein.

\bibitem{Bardeen}
J. Bardeen,
presented at GR5, Tiflis, U.S.S.R.,
and published in the conference proceedings in the U.S.S.R. (1968).

\bibitem{Zhou-Diss}
Z. Zhou,
Helv. Phys. Acta {\bf 65}, 767 (1992)
and references therein.

\bibitem{SGP}
H. Salazar, A. Garc\'{\i}a, and J. Pleba\'nski,
J. Math. Phys. {\bf 28}, 2171 (1987).

\bibitem{GR}
G.W. Gibbons and D.A. Rasheed,
Nucl. Pyhs. B {\bf 454}, 185 (1995).

\bibitem{Br-Lett}
K.A. Bronnikov,
Phys. Rev. Lett. {\bf 85}, 4641 (2000).

\bibitem{S-Diss}
O. Sarbach,
{\it On the generalization of the Regge-Wheeler equation for self-gravitating matter fields},
PhD thesis (2000), unpublished.

\bibitem{ST}
O. Sarbach and M. Tiglio,
Phys. Rev. D {\bf 64}, 084016 (2001).

\bibitem{Moncrief}
V. Moncrief,
Phys. Rev. D {\bf 9}, 2707 (1974);
Phys. Rev. D {\bf 10}, 1057 (1974);
Phys. Rev. D {\bf 12}, 1526 (1975).

\bibitem{C-Book}
S. Chandrasekhar,
{\em The Mathematical Theory of Black Holes}
(Oxford University Press, New York, 1983).

\bibitem{GS}
U.H. Gerlach and U.K. Sengupta,
Phys. Rev. D {\bf 19}, 2268 (1979).

\bibitem{Wald}
R.M. Wald,
J. Math. Phys. {\bf 33}, 248 (1992).

\bibitem{VG}
M.S. Volkov and D.V. Gal'tsov,
Phys. Lett. B {\bf 341}, 279 (1995).

\bibitem{PI}
E. Poisson and W. Israel,
Phys. Rev. D {\bf 41}, 1796 (1990).

\bibitem{ABMS}
E. Ay\'on-Beato, C. Moreno, and O. Sarbach,
in preparation.


\end{thebibliography}
\end{document}